\documentclass[twocolumn,showpacs,amsmath,amssymb]{revtex4}

\usepackage{graphicx}
\usepackage{dcolumn}
\usepackage{bm}
\usepackage{amssymb}

\begin{document}

\title{A quantum description of bubble growth in a superheated fluid}
\author{S. Choi$^{1}$, K. M. Galdamez$^{2}$, B. Sundaram$^{1}$}
\affiliation{$^1$ Department of Physics, University of
Massachusetts, Boston, MA 02125, USA \\
$^2$ Department of Physics and Astronomy, Tufts University, Medford,
MA 02155 USA}

\begin{abstract}
We discuss a quantum description of bubble growth in a superheated
liquid Helium by addressing the problem of operator ordering
ambiguities that arise due to the presence of position dependent
mass (PDM) in this system. Using a supersymmetric quantum mechanics
formalism along with the Weyl quantization rule, we are able to
identify specific operator orderings for this problem. This is a
general method which should be applicable to other PDM  systems.
\end{abstract}

\maketitle

\section{Introduction}

The study of bubble nucleation and growth is a broad field of study
with rich physics\cite{qnuc}. Some of the more recent examples
include experimental and theoretical studies of superfluid Helium in
a superheated state\cite{qnuc1}, and the theoretical study of
nucleation of magnetic bubbles\cite{qnuc2}. In this paper we look at
the problem of the quantum description of bubble growth in
superheated liquid Helium.  This is just one example of many systems
whose classical nature is well known but whose quantum
representation is not. The fact that not all functions can be
quantized uniquely into operators in the Hilbert space has been
known for some time. The Groenewald-van Hove theorem has explicitly
shown that Dirac's program of assigning operators to all functions
by having the Poisson brackets change into commutators works only
for polynomials of quadratic form or less\cite{ST}.

In the case of superheated liquid Helium, the difficulty in
quantization appears in the form of a position dependent mass (PDM),
whereby the kinetic term in the classical Hamiltonian is space
dependent. In fact, many systems in nature have the PDM structure;
for instance, free carriers, such as electrons, in semiconductors of
non-uniform chemical composition are often described by a PDM
Hamiltonian\cite{SC,SCPDM}. Other examples of PDM appear in various
instances of nuclear many-body problems, quantum dots, and
nano-mechanical systems just to name a few\cite{PDMexamples}. As a
result there have been a number of attempts to address the issue of
operator ordering ambiguities due to PDM, using the Gallileian
transformation and other methods\cite{PDMtheories}.

We aim in the present paper to provide insight into several existing
techniques in quantization, their convergence and possible future
usage. In particular, we show in this paper how to apply the
Supersymmetric quantum mechanics (SUSY QM) formalism\cite{Cooper}
which have been previously applied to solve simpler PDM
problems\cite{Cruz,Milanovic,Plastino}. The main difference is that,
unlike the usual examples of SUSY QM, our Hamiltonian contains a
specific potential which is not easily factorizable; the method
presented in this paper shows how one may apply SUSY QM to a more
general situation.

On the other hand, we apply to our system a more formal quantization
procedure which takes the Heisenberg commutation relations as
primary, the Weyl quantization otherwise referred to as the ``Weyl
Transform''\cite{ST,WQ,Cohen}. This allows one to quantize a PDM
Hamiltonian but does not provide explicit operator ordering. By
combining the two approaches namely the SUSY QM method and the Weyl
Transform, we are able to identify specifically a couple of correct
operator orderings out of infinitely many possible orderings.

The paper is divided as follows:  In Section 2, we give a brief
description of nucleation in superheated liquid Helium, which makes
the problem of PDM obvious. In Section 3, we describe the details of
using the SUSY QM formalism, to address the PDM problem. In Section
4, we introduce Weyl quantization and use it to obtain an operator
form for our classical Hamiltonian. This resulting Weyl quantum
Hamiltonian is converted to a point mass representation to compare
with the Hamiltonian obtained through SUSY methods. In Section 5 we
study the effective potential using realistic parameters, and
provide a conclusion.

\section{Bubble nucleation in Superheated Liquid Helium}

A superheated (or supercooled) fluid\cite{supercool} is typically
metastable since it cannot make a direct, uniform transition to the
stable phase throughout its volume. The transition to the stable
phase occurs via the nucleation of a droplet of stable fluid, such
that once the nucleus reaches a critical size it grows quickly,
ultimately converting the entire liquid from metastable to stable
phase. Nucleation can occur due to fluctuations in pressure and
volume within a liquid,  whereby small quantities of a new phase
(``bubbles'') form within a homogeneous phase.  If the liquid were
stable these bubbles would disappear; but superheated liquid Helium
is metastable, so these bubbles can grow and eventually transform
the system from the old liquid phase onto the new vapor phase.

This growth process is driven by pressure gradients across the surface of
the bubble. A recent study\cite{Gunther} provided a comprehensive
classical description of such a bubble nucleation process that takes
into account the fluctuations in both the radius and the pressure.
 The  Helmholtz free energy corresponding to this
process is:
\begin{eqnarray}
\Delta F  & = & \frac{2 \pi}{3} R^3 \kappa_1 (P_i-P_{v})^2 -
(P-P_{v}) \frac{4 \pi R^3}{3} \nonumber \\
& &  + 4 \pi \sigma R^2 \label{FreeEnergy}
\end{eqnarray}
where $R$ is the radius of the bubble and $P$ the pressure  outside of the bubble.
$P_i = P +  2 \sigma/R$
is the  pressure inside the bubble where $\sigma$ is the surface
tension,  $P_v$ is the internal equilibrium pressure of the bubble,
and $\kappa_i$ is the isothermal compressibility of the stable phase
at the equilibrium pressure. The first term
 does not affect the rate of nucleation since
it describes  the pressure fluctuations which average out to a
constant. The second and third terms of Eq. (\ref{FreeEnergy})
provide the barrier, with respect to the radius, over which the
nucleus must pass in order to expand and fill the volume with the
stable phase. The critical radius for vapor nucleation is then
$R_c =  2 \sigma/(P_v - P)$.

To establish the (classical) Hamiltonian for bubble nucleation, we note that the
kinetic energy of a growing nucleus is given by\cite{qnuc}
\begin{equation}
E_{k} =
\frac{1}{2}M(R) \left ( \frac{dR}{dt} \right )^2
\end{equation}
 where the variable mass is given by:
\begin{equation}
M(R) = 4 \pi \left ( 1 - \frac{\rho_v}{\rho_L} \right )^{2}
\rho_L R^{3}
\end{equation}
 with $\rho_L$ being the density of the liquid and
$\rho_v \ll \rho_L$ is the vapor density. The potential term based
on Eq. (\ref{FreeEnergy}) leads to the Hamiltonian for the bubble
nucleation problem \cite{Herbert}
\begin{equation}
H_{Class.} = \frac{p^2}{2M_0 x^3} + U_0 x^2(1-x) ,
 \label{HC}
\end{equation}
where  $x = R/R_c$ is the radius of the bubble scaled to the
critical radius and
$U_0  = 4 \pi\sigma R_c^2$  and $M_0 = 4 \pi \rho_L R_c^3$.
Returning to our particular case in question, we note that if we
were to quantize our classical PDM Hamiltonian naively by replacing
$x$ and $p$ by the corresponding position and momentum operators
$\hat{x}$ and $\hat{p}$, an infinite set of operator forms are
possible. For example, $\frac{p^2}{x^3} \rightarrow
\hat{p}\frac{1}{\hat{x}^3}\hat{p}$, $\frac{p^2}{x^3} \rightarrow
\frac{1}{2} \left [\hat{p}^2 \frac{1}{\hat{x}^3} +
\frac{1}{\hat{x}^3} \hat{p}^2 \right ]$, $\frac{p^2}{x^3}
\rightarrow \frac{1}{\hat{x}}\hat{p}
\frac{1}{\hat{x}}\hat{p}\frac{1}{\hat{x}}, \ldots$, to list just
some of the simplest cases; we show how to identify the correct
specific operator ordering below.

\section{Application of SUSY to PDM of non-symmetric potential}

While our potential is not super-symmetric, the supersymmetric
quantum mechanics (SUSY QM) formalism permits the re-writing of the
Hamiltonian in terms of the familiar creation-annihilation operators
by introducing an effective potential (``superpotential''). The
kinetic part of of a general PDM Hamiltonian (with $\hbar = 1$, and
without any confining potential) can be written
\begin{equation}
H  = - \frac{1}{2} {m(x)^a} \frac{d}{dx}  {m(x)^{2b}} \frac{d}{dx}  {m(x)^a},
\label{PDM}
\end{equation}
where $a$ (and $b$) can take any value as long as $a + b = -
\frac{1}{2}$. One can then write the creation-annihilation operators
$A_{a}^{\pm}$ as
\begin{eqnarray}
A_{a}^{-} & = &   \frac{1}{\sqrt{2}} {m(x)^b} \frac{d}{dx}  {m(x)^a}  +
W_{a}(x)  \label{Am} \\
A_{a}^{+} &  = &  - \frac{1}{\sqrt{2}} {m(x)^a} \frac{d}{dx} {m(x)^b}  +
W_{a}(x)    \label{Ap}
\end{eqnarray}
where $W_{a}(x)$ is known as superpotential and the corresponding
harmonic Hamiltonian is given by
\begin{equation}
H_{a}^{\pm} =  A_{a}^{\pm} A_{a}^{\mp} = T_{a}^{\pm} + V_{a}^{\pm}
\end{equation}
where
\begin{eqnarray}
T^{(+)}&  =  & - \frac{1}{2} {m(x)^a} \frac{d}{dx} {m(x)^{2b}}
\frac{d}{dx}  {m(x)^a}  \label{Tplus} \\
T^{(-)}&  =  & - \frac{1}{2} {m(x)^b} \frac{d}{dx} {m(x)^{2a}}
\frac{d}{dx}  {m(x)^b}
\end{eqnarray}
are the kinetic terms and $V_{a}^{\pm}$ are the corresponding
potential terms that effectively play the role of the quadratic
confining potential of a standard harmonic oscillator. The form of
$W_a$ is fixed by demanding that $A_ {a}^{\pm}$ obey  the Heisenberg
algebra i.e. $[A_{a}^{-},A_{a}^{+}] = 1$. This condition leads to
\begin{eqnarray}
W_a & = & \frac{1}{2} \int \sqrt{2 m(x)} dx + \frac{4a + 1}{2}
 \left ( \frac{1}{\sqrt{2 m(x)}} \right )'  ,
\end{eqnarray}
which then implies
\begin{eqnarray}
V_{a}^{\pm} & = &  \frac{1}{2} \left ( \int \sqrt{m(x)} dx  \right
)^2 + \frac{4a + 1}{4} \frac{1}{\sqrt{m(x)}} \left (
\frac{1}{\sqrt{m(x)}} \right )'' \nonumber \\
& - & \frac{(4a + 1)^2}{8} \left ( \frac{d}{dx}
\frac{1}{\sqrt{m(x)}} \right )^2 \mp \frac{1}{2} .
\end{eqnarray}
We emphasize here that the procedure is independent of the form of
the mass -- the mass $m(x)$ is completely general. The essence of
the SUSY procedure is therefore the introduction of an extra
potential $V_{a}$ which now contains all the effect of operator
ordering via the parameter $a$. For our specific PDM corresponding
to the case of superheated liquid Helium, $m(x) = M_0 x^3$ so that:
\begin{equation}
 W_a   =  -\frac{3(4a + 1)}{\sqrt{32  M_0 x^5}} +
\frac{1}{5}\sqrt{2
 M_0   x^{5}}  \label{Wa}
\end{equation}
and
\begin{equation}
V_{a}^{\pm}  =  \frac{21 + 48 a - 144 a^2}{32  M_0   x^5} +
\frac{2}{25}  M_0  x^{5} \mp \frac{1}{2} . \label{VaP}
\end{equation}

It is noted that when $a = b = -\frac{1}{4}$
\begin{equation}
W_a = \frac{1}{2} \int \sqrt{2 m(x)} dx ,  \;\;\;  {\rm and} \;\;\;
V_{a}^{\pm}  =
 W_a^2  \mp \frac{1}{2},
 \end{equation}
 which are identical, respectively, to the
equivalent superpotential and potential obtained in the classical
formalism using Poisson brackets in place of commutators\cite{Cruz}.

To solve our problem, we need to write the Hamiltonian corresponding
to our situation in terms of the operators $A_{a}^{\pm}$. Writing
out the ``momentum'' part of $A_{a}^{\mp}$ using our PDM $m(x) = M_0
x^3$, one can show that $ A_{a}^{+} + A_{a}^{-}  = \frac{2}{5}
\sqrt{2  M_0  x^{5} }$
 which does not depend on the parameter $a$ i.e independent of ordering.
Using this result, one is able to write down the
 potentials $V_{a}^{\pm}$ and $U_0 x^2(1-x)$ in terms of $A_{a}^{+} + A_{a}^{-}$
 in a straightforward manner.
In terms of these operators, the Hamiltonian for our system is given
by subtracting $V_{a}^{+}$ from  $A_{a}^{+}A_{a}^{-}$  and building
the $U_0 x^2(1-x)$ potential by remembering that $x$ is unitless in
the original Hamiltonian -- this is equivalent in this case to
multiplying our $x$ by a factor $(8 M_0/25)^{1/5}$ to get:
\begin{eqnarray}
{H}_{ Quant.} & = & A_{a}^{+}A_{a}^{-} - \frac{21 + 48 a - 144
a^2}{100} (A_{a}^{+} + A_{a}^{-})^{-2} \nonumber \\
& &  -  \frac{1}{4}(A_{a}^{+} + A_{a}^{-})^2  + U_0 (A_{a}^{+} + A_{a}^{-})^{4/5}  \nonumber \\
 & \times & \left [ 1  -  (A_{a}^{+} + A_{a}^{-})^{2/5} \right
 ] + \frac{1}{2} .  \label{HQ}
\end{eqnarray}
The Hamiltonian of Eq. (\ref{HQ}) is our PDM Hamiltonian when
$A_{a}^{\pm}$ is defined as in Eqs. (\ref{Am}-\ref{Ap}) with the
superpotential $W_a$ of Eq. (\ref{Wa}). Effectively, we have
re-written our Hamiltonian solely in terms of the new raising and
lowering operators $A_{a}^{+}$ and $A_{a}^{-}$. We note again that
in Eq. (\ref{HQ}), the kinetic energy term is still implicitly with
an indeterminate operator ordering i.e. $A_{a}^{+}A_{a}^{-} -
V_{a}^{+} = T^{(+)}$ of Eq. (\ref{Tplus}).


We note that the operators $A_{a}^+$ and $A_{a}^-$ function as
raising and lowering operators and are constructed such that they
abide to the Heisenberg algebra, $[A_{a}^-, A_{a}^+]=1$,  in the
same way as the well-known raising and lowering operators for the
standard quantum harmonic oscillator (QHO). Guided by the
definitions of the standard QHO, we therefore create an ansatz
whereby we define new ``position'' and ``momentum'' variables: $x' =
\sqrt{ \frac{\hbar} {2M_0 \omega}} (A_{a}^{+} + A_{a}^{-})$ and $p'
= i\sqrt{ \frac{\hbar M_0 \omega} {2}} (A_{a}^+ - A_{a}^-)$. $x'$
and $p'$ then play the role of position and momentum operators with
the usual commutation relation, $[x', p'] = i$. It is noted that the
commutation relation for $x'$ and $p'$ implies $p'= i \frac{d}{dx}$
via the Stone-von Neumann theorem\cite{SvN} which states that given
$[\hat{a}, \hat{b}] = i$, there is a representation such that
$\hat{a}| \psi \rangle = x | \psi \rangle$ and $\hat{p}|\psi \rangle
=  i \frac{d}{dx} | \psi \rangle$. In these definitions, $\omega$
and $M_0$ correspond to the mode energy and the mass of the bubble.
Scaling our Hamiltonian by multiplying throughout by $\hbar \omega$
and changing the variable to $z \equiv x'/R_c$ for a later
comparison (note that the variable $z$ in this section relates to
$x'$. Here $z$ is {\it not} a coordinate transform of the original
position variable $x$ as done in the next section), we are led to
the following Hamiltonian by a straightforward substitution of $x'$
and $p'$ into Eq. (\ref{HQ}):
\begin{equation}
H_{eff.} =  -\frac{\hbar^2}{2M_0 R_c^2} \frac{d^2}{dz^{2}} +
V_{a}(z)  + V_{sys}(z) \label{Heff}
\end{equation}
where
\begin{equation}
 V_{a}(z) =  -\frac{\hbar^2}{2M_0 R_c^2} \left [ \frac{21 + 48a - 144 a^2}{100
 z^{2}} \right ]
\end{equation}
and
\begin{equation}
V_{sys}(z) =  U_0  z^{4/5} \left [ 1 -z^{2/5} \right ] + \frac{1}{2} .
\end{equation}
Here $U_0$ is modified to take into account the scaling of
variables. The effective potential, $V_{a}(z)+ V_{sys}(z)$, is
fairly complicated compared to the original potential; in particular
it contains a singularity at $z = 0$ -- such is the ``price'' of
removing the PDM from the kinetic term. The energy spectrum of the
effective Hamiltonian of Eq. (\ref{Heff}) is identical to that of
the original PDM Hamiltonian (up to a constant factor). On the other
hand, because the position variable $z$ is now in a different space,
the eigenstate of Eq. (\ref{Heff}) is not the eigenstate of the
original Hamiltonian, although there are ways to map the eigenstate
of Eq. (\ref{Heff}) back onto the eigenstate of the original PDM
Hamiltonian. Once the parameter $a$ is determined, the unique
operator ordering is decided; we now suggest a way to select the
value of this parameter.

We note here that the steps involved to derive Eq. (\ref{Heff}) is
that of  ``picking a particular basis'' for the Hamiltonian of Eq.
(\ref{HQ}) rather than a coordinate transformation from one position
coordinate system to another position coordinate system. In
particular, by choosing ansatz based on the well-known quantum
harmonic oscillator, the resulting Hamiltonian is assured of being
of unit measure and Hermitian. These requirements of unit measure
and Hermiticity is explained in more detail in the next section,
where a coordinate transformation is directly involved.

\section{Comparison with Weyl Quantization}

There exist in the literature several ways to formally convert
classical function into operator form. Weyl quantization otherwise
referred to as the ``Weyl Transform'' is a well-known way of doing
so. There are many qualities of the Weyl transform that make it an
optimum way to perform this conversion\cite{ST,WQ,Cohen}.
The Weyl transform is defined as:
\begin{eqnarray}
W[FT \phi](x) & =&  \frac{1}{2 \pi} \int \int [FT](\alpha, \beta)
e^{i\hbar \alpha \beta/2} e^{i \beta x} \nonumber \\
& & \times  \phi(x + \hbar \alpha) \Omega(\alpha,\beta) \/ d\alpha
\/ d \beta . \label{WT}
\end{eqnarray}
Here
\begin{equation}
[FT](\alpha, \beta)= \int T(p,x) \exp [-i(p \alpha + x \beta)] \/ dp
\/ dx
\end{equation}
denotes the Fourier transform of the classical function, $T(x,p)$
and $\phi(x)$ is the wave function upon which the Weyl operator
acts.  The weighting factor, $\Omega(\alpha, \beta)$, characterizes
the type of quantization to be utilized.  Weyl quantization, in
particular, is defined when $\Omega(\alpha, \beta) = 1$.  Therefore,
the Weyl transform of the kinetic part of our classical Hamiltonian
is
\begin{eqnarray}
 && W \left [  F \left (\frac{p^2}{2 M_0 x^3} \right ) \phi \right ](x) =
\frac{1}{ 2 \pi}
 \int \int F \left ( \frac{p^2}{2 M_0 x^3} \right ) \nonumber \\
&& \times e ^{i \hbar \alpha \beta/2} e^{i  \beta x} \phi (x + \hbar
\alpha) \/ d\alpha \/ d\beta .
\end{eqnarray}
Using the result
\begin{equation}
F \left ( \frac{p^2}{ x^3} \right ) =  [ - \sqrt{2 \pi}
\delta''(\alpha)] \left [ -i \sqrt{ \frac{\pi}{2 }} \left ( -
\frac{\beta^2}{2} sgn \beta \right ) \right ] ,
\end{equation}
the Weyl transform for our case is
\begin{eqnarray}
&& W \left [  F \left (\frac{p^2}{2 M_0 x^3} \right )  \phi \right ]
 =  - \frac{\hbar^2}{2 M_0 R_c^2}  \nonumber \\
&&  \times  \left [ \frac{1}{x^3} \frac{\partial^2}{\partial
x^2}\phi(x) - \right.
 \left. \frac{3}{x^4}\frac{\partial}{\partial x} \phi(x) +
\frac{3}{x^5} \phi(x) \right ] . \label{WeylEq.}
\end{eqnarray}
We note here that the Hermiticity of the Hamiltonian must be
ensured. One can show that, given a general Hamiltonian of the form
${\hat{H}} = {A}(x) \frac{\partial^{2}}{ \partial x^{2}} + {B}(x)
\frac{\partial}{\partial x} + {C}(x)$ the Hermiticity condition,
\begin{equation}
\int {\psi^{*}} {\hat{H}} {\phi}(x) dx = \int \left [ {\hat{H}}
{\psi}(x) \right ]^{*} {\phi}(x) dx \label{herm}
\end{equation}
imposes the following conditions after integrating various terms by
parts: (i) $A^{*}(x) = A(x)$ i.e. $A(x)$ is real; (ii)  $A'(x) =
{\rm Re}[B(x)]$; (iii) ${\rm Im}[C(x)] = \frac{1}{2} {\rm
Im}[B'(x)]$ where ${\rm Re}[\cdot]$ and ${\rm Im}[\cdot]$ denote
real and imaginary parts respectively. This general result shows
that the Weyl transform such as the result of Eq. (\ref{WeylEq.})
preserves the Hermiticity of the Hamiltonian.

In order to compare this result with the Hamiltonian obtained in
SUSY, we begin with Eq. (\ref{WeylEq.}) and note that the associated
kinetic term in the corresponding Lagrangian is
\begin{equation}
L_K(x, \dot{x}, t) = \frac{1}{2}M_0 R_c^2 x^3 \dot{x}^2 .
\label{LPDM}
\end{equation}
We introduce a coordinate transformation $x \rightarrow c
z^{\alpha}$ where $c$ and $\alpha$ are real numbers transforming
$L_{K}(x, \dot{x}, t)$  to  $L_{K}(z, \dot{z}, t) = \frac{1}{2}M_0
R_c^2  c^{5}  \alpha^2 z^{5 \alpha - 2} \dot{z}^2$. In order for the
$1/x^3$ factor in the first term to disappear, one needs to choose $\alpha =
\frac{2}{5}$ and $c = \left (\frac{5}{2} \right )^{2/5}$ i.e.  $x
\rightarrow \left (\frac{5z}{2} \right )^{2/5}$, so that the mass
term becomes a constant, $m(x) \rightarrow 1$ and Eq. (\ref{LPDM})
is  transformed to
\begin{equation}
L_K(z, \dot{z}, t) =  \frac{1}{2}M_0 R_c^2  \dot{z}^2.
\end{equation}
With this substitution, Eq.(\ref{WeylEq.}) is transformed into
\begin{equation}
\hat{H }_{K}(z)= -\frac{\hbar^2} {2 M_0 R_c^2} \left [
\frac{\partial^2}{\partial z^2}  - \frac{3}{5 z}
\frac{\partial}{\partial z} + \frac{12}{25 z^2} \right ]. \label{PMR}
\end{equation}

Once the coordinate transform is carried out, the measure (or the
Jacobian of integration $|\frac{dz}{dx}|$) associated with the
Hamiltonian of Eq.(\ref{PMR}) is no longer unity i.e. coordinate
transformation does not preserve the inner product of the Hilbert
space, which is problematic.  In general, unit measure can be
restored through the comparison of the inner products between two
Hamiltonians $\tilde{\hat{H}}$ and $\hat{H}$ where $\tilde{\hat{H}}$
is the Hamiltonian for the inner-product space with unit measure and
$\hat{H}$ corresponds to that with non-unit measure $\mu(x)$ so as
to preserve the Hilbert space:
\begin{equation}
\int \psi^{*}(z) \hat{H} \phi(z) \mu(z) dz = \int
\tilde{\psi}^{*}(z) \tilde{ \hat{H} } \tilde{\phi}(z) dz.
\label{meas}
\end{equation}
Noting that $f(z) = \tilde{f}(z)/\sqrt{\mu(z)}$ where $f(z) =
\phi(z)$ or $\psi(z)$, Eq. (\ref{meas}) implies the relationship
between the two Hamiltonians as $( \tilde{\hat{H}} \bullet) = \left
( \sqrt{\mu(z)} \hat{H} \frac{1}{\sqrt{\mu(z)}} \bullet \right )$.
Writing a general unit measure Hamiltonian  $\tilde{\hat{H}}$ as
\begin{equation}
\tilde{\hat{H}} = \tilde{A}(z) \frac{\partial^{2}}{ \partial z^{2}}
+ \tilde{B}(z) \frac{\partial}{\partial z} + \tilde{C}(z) ,
\label{unitH}
\end{equation}
and the non-unit measure Hamiltonian as one without the tildes, the
above relation between $\hat{\tilde{H}}$ and $\hat{H}$ implies
\begin{eqnarray}
\tilde{A}(z)& = & A(z) , \\
\tilde{B}(z) & = & B(z) - A(z) \frac{\mu'(z)}{\mu(z)} , \\
\tilde{C}(z) & = & A(z) \left [ \frac{3}{4}
\frac{\mu'(z)^2}{\mu(z)^{2}} - \frac{1}{2} \frac{\mu''(z)}{\mu(z)}
\right ] \nonumber \\
& &  -  B(z) \left [ \frac{1}{2} \frac{\mu'(z)}{\mu(z)} \right ]+
C(z) .
\end{eqnarray}
We note that these results are general results for any Hamiltonian
of the form Eq. (\ref{unitH}); it is not a property of a specific
Hamiltonian, or any transforms involving differential operators, but
is rather a self-consistent result following from preserving the
inner product, and ensuring Hermiticity.  A corollary of these
results is that for a non unit measure Hamiltonian, the Hermiticity
condition implies $B(z) = A'(z) + \frac{ \mu'(z)}{\mu(z)}A(z)$ i.e.
the Hamiltonian of Eq. (\ref{PMR}) is still Hermitian. This relation
also means that a point-mass, unit measure Hamiltonian will always
have $\tilde{B}(z) = 0$ due to Hermiticity condition  i.e. the first
order derivative term always vanishes, which is what we normally see
in ``ordinary'' Hamiltonians. In our case, since $x = \left (
\frac{5}{2}z \right )^{2/5}$, the measure is calculated to be
$\mu(z) = \frac{dz}{dx} = \left ( \frac{2}{5z} \right )^{3/5}$. The
Hamiltonian with unit measure emanating from Eq.(\ref{PMR}) is then
\begin{equation}
\hat{H}(z) = -\frac{\hbar^2} {2 M_0 R_c^2 } \left [ \frac{d^2}{dz^2}
+ \frac{9}{100 z^2} \right ] +  U_0 z^{4/5} (1 - z^{2/5}) .
\label{HQz}
\end{equation}
Equating this result of Eq. (\ref{HQz}) with the Hamiltonian
obtained through SUSY QM, Eq. (\ref{Heff}), we require
 \begin{equation}
 \frac{9}{100 z^2}  = \frac{21 + 48a - 144 a^2} {100
z^{2} },
\end{equation}
which implies that specific operator orderings that corresponds to
the Weyl quantization rule are given by
$a \rightarrow -\frac{1}{6} \;\;\;\; {\rm or} \;\;\;\; a \rightarrow
\frac{1}{2}$.
Therefore, the quantum Hamiltonian for the bubble nucleation problem
that obeys the Weyl quantization rule is
either
\begin{equation}
H  = - \frac{\hbar^2}{2 M_0 R_c^2} \left [ \frac{1}{x^{1/2}} \frac{d}{dx}
 \frac{1}{x^{2}} \frac{d}{dx}  \frac{1}{x^{1/2}} \right ]  + U_0 x^2(1-x)
\end{equation}
or
\begin{equation}
H  = - \frac{\hbar^2}{2 M_0 R_c^2} \left [ x^{3/2} \frac{d}{dx}
 \frac{1}{x^{6}} \frac{d}{dx}  x^{3/2} \right ]  + U_0 x^2(1-x)  .
\end{equation}
We have therefore narrowed down the possible operator orderings to
two (out of infinity). This is the central result of our paper.

We note here that it seems on first sight that one could avoid the
operator ordering ambiguity from the outset by transforming the
classical Hamiltonian to a point mass form first, and then
quantizing it. However, the proper procedure is quantization and
then coordinate transform as we have done in this paper. This is
because one loses more information than necessary otherwise -- the
extra information that a Weyl Transform gives us as a result of
having PDM is lost by carrying out the coordinate transformation
first. More specifically, one would not obtain the term proportional
to $1/z^2$ if the point mass transform was done first. This would
also be inconsistent with the SUSY procedure of Section 3 which
independently gave the term proportional to $1/z^2$.

Finally, we note the deeper connection between the methods of
Section 3 and this section. The fact that the relatively simple
procedure based on SUSY QM formalism matches the result of Section 4
that involves Weyl transform, coordinate transform for finding a
point mass representation and re-scaling to unit measure is, indeed,
a demonstration of mathematical self-consistency. At a deeper level,
the match is not surprising since the relationship between $x$ and
$x'$ of Section 3 is indeed the same as the coordinate transform of
Section 4, namely $x' = (2/5) x^{5/2}$. However, the method of
Section 3 is somewhat more efficient than that of Section 4 since it
bypasses the need to find the point mass representation and to
restore unit measure. This efficiency is possible since the
quantization takes place not in the position space but in the space
of raising and lowering operators; picking a useful position
representation related to well known quantum harmonic oscillator
then ensures Hermiticity and unit measure owing to the Stone-von
Neumann theorem.

\section{Concluding remarks}

The goal of this paper was in finding the correct operator ordering
for the bubble nucleation problem. A full quantum mechanical
treatment of the bubble nucleation process that addresses tunneling
through the barrier will be presented elsewhere\cite{micro}. In this
section, before we conclude the paper, we attempt to gain some
physical insight into this problem by briefly examining the
effective Hamiltonian of Eq. (\ref{Heff}) [and Eq. (\ref{HQz})] with
realistic parameters. We consider physically reasonable parameters
for this system so that we can get an order of magnitude estimate
for the parameters. The typical values for superfluid Helium at
temperature $T = 4K$ is $\sigma = 0.12 \times 10^{-3} N/m$, $P_v =
8.1445 \times 10^{4} N/m^2$, $\rho_L = 140 kg/m^3$. For zero applied
pressure, the classical critical radius is $R_c = 29.5 \times
10^{-10} m$. At $T=4K$, the thermal de Broglie wavelength of a
single Helium atom is $\Lambda = h/\sqrt{2\pi m k T} \approx 4.36
\times 10^{-10} m$, and hence thermal momentum of $p_{Th} =
h/\Lambda = 1.52 \times 10^{-24} kg ms^{-1} \gg \sqrt{U_0 M_0}$ over
all range of applied pressure $P$. Typically a single bubble
contains an order of 100 Helium atoms, or momentum of roughly $10
p_{Th}$.

\begin{figure}
\begin{center}
\centerline{\includegraphics[height=7cm]{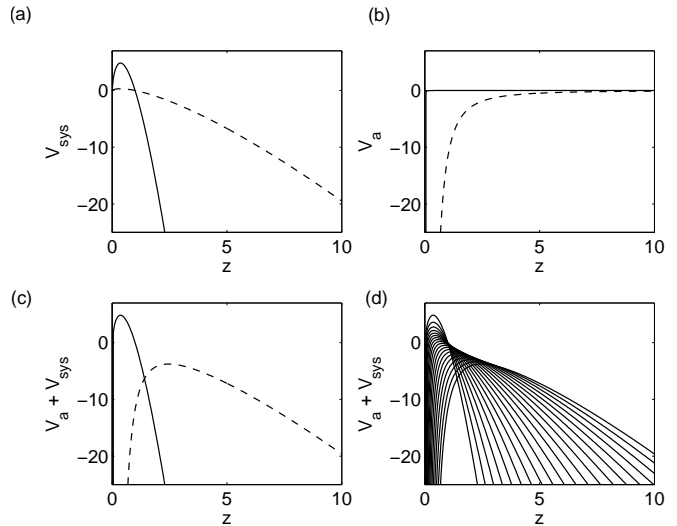}} \caption{(a)
$V_{sys}(z)$ for pressure $P = 0.8P_v$ (Dashed line) and $P =
0.95P_v$ (Solid line). (b) Same as (a) but for $V_{a}(z)$ instead.
(c) $V_{a}(z)+V_{sys}(z)$ for the same values of pressure. (d)
$V_{a}(z)+V_{sys}(z)$ for several values of pressure $P \in [0.8P_v,
0.95P_v]$. In this figure, the position $z$ is in scaled units (in
units of $R_c$) and the potential is in eV.} \label{Potential}
\end{center}
\vspace{-0.75cm}
\end{figure}

With these realistic numbers, we have calculated and plotted the
shape of $V_{a}(z) = -\frac{\hbar^2} {2 M_0 R_c^2 }  \frac{9}{100
z^2}$, and $V_{sys}(z) = U_0 z^{4/5} (1 - z^{2/5})$ in
Fig.~\ref{Potential} for various values of  applied pressure
$P/P_v$. The non-confining shape of the potential makes it clear why
the numerical solution to the eigenvalue problem is difficult and
non-convergent. The changing shape of $V_{a}(z)$ near $z = 0$ due to
changes in pressure and subsequently $R_c$ is also shown in Fig.
\ref{Potential}. On the other hand, the potential for larger $z$
corresponding to $V_{sys.}(z)$ looks simpler. It is clear from Fig.
\ref{Potential} demonstrating unbounded potential that the energy
spectrum is expected to be continuous, and the bubble growth is
similar to a ball rolling downhill with the ``gradient'' dependent
on the applied pressure. We note that had we not transformed the
Hamiltonian and left the PDM in the kinetic energy term we would not
be able to come up with such an intuitive understanding of the
system.

To conclude, we have studied the bubble nucleation problem in a
superheated liquid Helium and provided a quantum mechanical
description. Although the system seems rather complicated at the
outset and furthermore involves an exotic Hamiltonian that diverges
at the origin due to the specific form of the PDM, we were able to
extract useful information through the methods of SUSY QM and the
Weyl transform. We have presented a convergence between two methods
by converting to a point mass description of the quantum
Hamiltonian. Further analysis will need to be performed in regards
to comparison to the microscopic description of the problem at hand,
and these will be presented elsewhere\cite{micro}.


\begin{thebibliography}{00}

\bibitem{qnuc} I. M. Lifshitz and Yu. Kagan JETP {  35} (1972) 206

\bibitem{qnuc1} H. Maris, and S. Balibar Phys. Today {  53} (2000) 29

\bibitem{qnuc2}  E. M. Chudnovky and L. Gunther Phys. Rev. B {  37}
(1988) 9455 ; L. Gunther and A. DeFranzo, {\it ibid.} {  39} (1989)
11755

\bibitem{ST} V. Guillemin, S. Sternberg {\it Symplectic Techniques in Physics}
Cambridge University Press (1984)

\bibitem{SC} T. Gora and F. Williams, Phys. Rev. {177} (1969) 3

\bibitem{SCPDM} O. Roos, Phys. Rev. B {27} (1983) 12

\bibitem{PDMexamples} G. Bastard, {\it Wave mechanics applied to
Semiconductor Heterostructures} Editions de Physique, Les Ulis,
(1998);  P. Ring and P. Schuck, {\it The Nuclear Many body Problem}
Springer New York (1980); A. Dutra, C. A. S. Almeida, Phys. Lett. A.
{275}, (2000), 25; A. D. Alhaidari, Phys. Rev. A {66}, (2002), 66;
C.-Y. Cai , Z.-Z. Ren , and G.-X. Ju, Commun. Theor. Phys. {43},
(2005), 1019

\bibitem{PDMtheories} J.R. Shewell Am. J. Phys. {  27} (1959) 16; J.M. L\'{e}vy-Leblond Phys. Rev. A {  52} (1995) 1845
;  A. Ganguly, S. Kuru, J. Negro, L. M. Nieto, Phys. Lett. A { 360}
(2005) 228

\bibitem{Cooper} F. Cooper, A. Khare, U Sukhatme, Phys. Rep. A {  251} (1995) 267

\bibitem{Cruz} S. Cruz y Cruz, J. Negro, L. M. Nieto, Phys. Lett. A {  369} (2007) 400

\bibitem{Milanovic}  V. Milanovic and J. Ikovic, J. Phys. A {  32} (1999) 7001

\bibitem{Plastino}  A. R. Plastino, A. Rigo, M. Casas, F. Gracias, and A. Plastino
Phys. Rev. A {  60} (1999) 4318

\bibitem{WQ} D. A. Dubin {\it Mathematical Aspects of Weyl Quantization and Phase} World Scientific Publishing Co. Pte. Ltd.
(2000)

\bibitem{Cohen} L. Cohen, J. Math. Phys. {7}, (1966)  781

\bibitem{supercool} K. Huang {\it Statistical Mechanics} Wiley, New
York (1987); C. Kittel and H. Kroemer {\it Thermal Physics} W. H.
Freeman, San Francisco (1980)

\bibitem{Gunther}  L. Gunther Am. J. Phys. {  71} (2003) 351

\bibitem{Herbert} H. Simanjuntak  {\it Quantum nucleation of vapor in a metastable
liquid} (Unpublished)

\bibitem{SvN} H. Weyl, {\it The Theory of Groups and Quantum Mechanics}, Dover
Publications (1950)

\bibitem{micro} K. Galdamez (In preparation); M. Olshanii,  V.
Dunjko, S. Jackson (Private communication)



\end{thebibliography}
\end{document}